\documentclass[showpacs,twocolumn,showkeys,amsmath,amssymb,prr]{revtex4-1}

\usepackage{bm}
\usepackage{bbm}
\usepackage{dsfont}
\usepackage{color}
\usepackage{graphicx}  
 \usepackage[colorlinks=true,urlcolor=blue,linkcolor=green]{hyperref}

\newcommand{\be}{\begin{equation}}
\newcommand{\ee}{\end{equation}}
\newcommand{\rd}{{\mathrm d}}
\newcommand{\re}{{\mathrm e}}
\newcommand{\ri}{{\mathrm i}} 
\newcommand{\lF}{\langle\!\langle}
\newcommand{\rF}{\rangle\!\rangle}
\newcommand{\ab}{a^{\phantom{\dagger}}}
\newcommand{\ad}{a^{\dagger}}
\begin{document}

\title[Pre-Floquet states]
      {Pre-Floquet states facilitating coherent subharmonic response of 
      periodically driven many-body systems}  

\author{Steffen Seligmann, Hamed Koochaki Kelardeh, and Martin Holthaus}
		
\affiliation{Institut f\"ur Physik, Carl von Ossietzky Universit\"at,
	D-26111 Oldenburg, Germany}	
                  
\date{August 29, 2025}

\begin{abstract}
We demonstrate longtime coherent subharmonic motion of a  
many-boson system subjected  to an external time-periodic driving force. 
The underlying mechanism is exemplified numerically through analysis of 
a periodically driven Bose-Hubbard dimer, and clarified conceptually by 
semiclassical requantization of invariant tubes pertaining to the system's 
mean-field description. In this way, one arrives at pre-Floquet 
states that relate to the actual many-body Floquet states in a manner similar 
to the relation of site-localized Wannier states to lattice-extended Bloch 
states in solid-state physics.   It is argued that even high-order subharmonic 
response can be systematically engineered, and be observed experimentally, 
with weakly interacting Floquet condensates comprising a sufficiently large 
number of particles.   
\end{abstract} 


\maketitle 


\section{Introduction: Time crystals and subharmonic response}
\label{S_1}

While the thought-provoking question whether {\em continuous\/} 
time-translation symmetry could be spontaneously broken in the ground state of 
a closed quantum mechanical system~\cite{Wilczek12} soon was given a negative 
answer~\cite{Bruno13,WatanabeOshikawa15}, breaking of the {\em discrete\/} 
time-translational symmetry inherent to systems exposed to an external
time-periodic stimulus has actually been observed in pioneering experiments
with interacting spin chains of trapped atomic ions~\cite{ZhangEtAl17}, or 
with dipolar spin impurities in diamond~\cite{ChoiEtAl17}. The hallmark of 
such discrete time crystals is a subharmonic response to the periodic drive. 
Considering, for instance, $1:2$~clocking, that is, a response signal that 
occurs with strict periodicity only once during every two drive cycles, that 
signal could show up either in the first or in the second cycle of each 
two-cycle interval. This leaves us with two possible states, akin to the two 
ground states located in either well of a symmetric double-well potential 
when the tunneling contact between the two wells is closed. The documented 
existence of this nonequilibrium state of matter has catalyzed a multitude of 
further intense research, spanning, among others,  Anderson localization in 
the time domain, ergodicity breaking, and lack of thermalization due to 
many-body localization, altogether disclosing far-reaching new perspectives 
for nonequilibrium statistical physics~\cite{KeyserlingkEtAl16,ElseEtAl16,
RussomannoEtAl17,SachaZakrzewski18,SuraceEtAl19,KhemaniEtAl19,
GuoLiang20,ElseEtAl20,PizziEtAl21,ZalatelEtAl23}.

The purpose of the present contribution is to specify conditions under which 
long-lasting subharmonic response of many-body quantum systems 
occurs in an ele\-mentary manner not involving these demanding concepts. 
For the sake of demonstration, we resort to the model of a periodically driven 
Bose-Hubbard dimer, describing a large number~$N$ of Bose particles that 
occupy two sites coupled by a tunneling matrix element~$\hbar\Omega$, and 
experience a repulsive on-site interaction of strength~$\hbar\kappa$. Instead 
of periodic modulation of the tunneling contact~\cite{KiddEtAl20} or delta-like 
kicking~\cite{LiangEtAl24},  here we  consider sinusoidal shaking of the site 
potentials with  amplitude~$\hbar\mu$ and angular frequency~$\omega$. In 
terms of bosonic operators $\ab_j$ and $\ad_j$ that annihilate and create, 
respectively, a particle at the site labeled~$j$, its Hamiltonian takes the 
form~\cite{HolthausStenholm01,WeissTeichmann08} 
\begin{eqnarray}
	H(t) & = & -\frac{\hbar\Omega}{2} 
	\left( \ad_2\ab_1 + \ad_1\ab_2 \right) 
\nonumber \\ & & 
	+ \hbar\kappa \left( \ad_1\ad_1\ab_1\ab_1 
	+ \ad_2\ad_2\ab_2\ab_2 \right) \
\nonumber \\ & &	
	+ \hbar\mu \sin(\omega t)
	\left( \ad_1\ab_1 - \ad_2\ab_2 \right) \; .  
\label{eq:DBH}		
\end{eqnarray}
This model~(\ref{eq:DBH}) is closely related to the driven 
one-dimensional spin chain with long-range interactions more recently studied 
in Ref.~\cite{PizziEtAl21}.

\begin{figure}[t]
\centering
\includegraphics[width=1.0\linewidth]{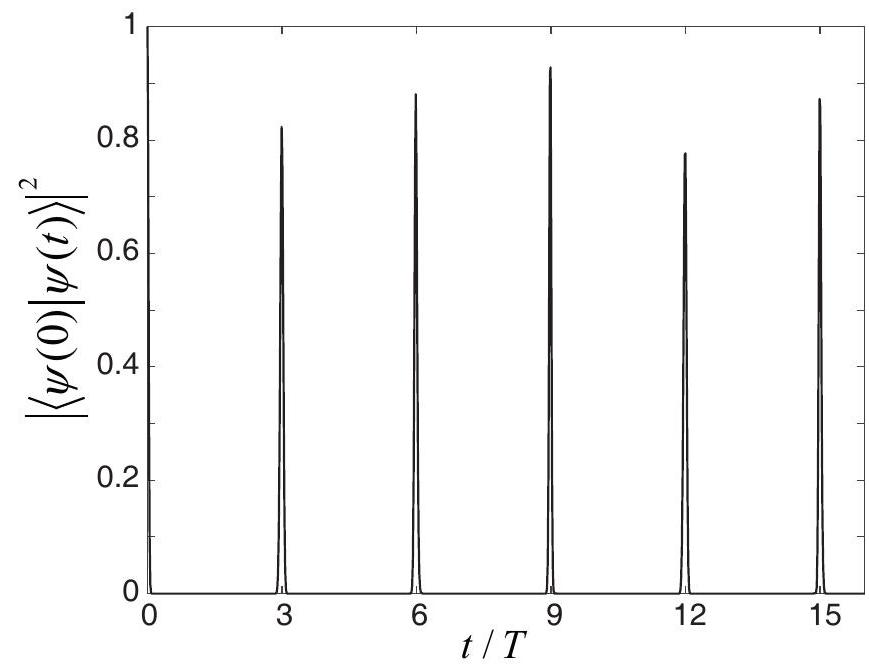}
\caption{Return probability 
$P_r(t) = \left| \langle \psi(0) | \psi(t) \right|^2$
for a state $|\psi(t)\rangle$ of the periodically driven Bose-Hubbard
dimer~(\ref{eq:DBH}) with $N = 2000$ particles. Here, the time~$t$ is scaled 
with respect to the cycle duration~$T = 2\pi/\omega$, revealing $1:3$ 
subharmonic clocking. Dimensionless system parameters are 
$N\kappa/\Omega = 0.92$, $\mu/\Omega = 0.4$, and 
$\omega/\Omega = 1.9$.}
\label{F_1}
\end{figure}

Having solved the system's Schr\"odinger equation numerically with $N = 2000$ 
particles in order to obtain the time-dependent states $|\psi(t)\rangle$,  
we depict in Fig.~\ref{F_1} the return probability 
$P_r(t) = \left| \langle \psi(0) | \psi(t) \right|^2$ 
vs.\ time~$t$ to a particular initial state $|\psi(0)\rangle$ under conditions 
of strong driving, $\mu/\Omega = 0.4$ for $\omega/\Omega = 1.9$ and 
$N\kappa/\Omega = 0.92$. Evidently, the system features almost perfect $1:3$
clocking, at least on the time scale of few driving cycles considered here. 
In the following sections, we will analyze the underlying mechanism 
that may be considerably more general than its realization in the idealistic 
model~(\ref{eq:DBH}). To this end, we will review key attributes of 
the Floquet approach to periodically driven quantum systems in Sec.~\ref{S_2},
emphasizing the close connection between Sambe's extended Hilbert space 
and the Brillouin-zone-like structure of the quasienergy spectrum. In 
Sec.~\ref{S_3}, we outline how this structure is recovered by a generalization 
of the well-known Bohr-Sommerfeld quantization rules to fully integrable, 
periodically driven classical Hamiltonian systems. In the central 
Sec.~\ref{S_4}, we combine these two strands, applying the quantization rules 
to almost inte\-grable mean-field motion prevailing in resonance zones 
surrounded by a chaotic sea. This leads to the concept of pre-Floquet states 
with a period other than the driving period that are capable of sustaining 
longtime subharmonic quantum evolution. The final Sec.~\ref{S_5} suggests 
the possibility to engineer high-order subharmonic response experimentally, 
employing Floquet condensates.  In the Appendix, we argue that there exists an 
inevitable residual hybridization of regular and  chaotic Floquet states, 
effectuating gradual dephasing of coherent subharmonic dynamics.

\section{Preliminaries: Historical development of the Floquet picture 
and some of its subtleties}
\label{S_2}

As is common practice by now, the investigation of the quantum dynamics 
generated by some periodically time-dependent Hamiltonian $H(t) = H(t+T)$ on 
its Hilbert space~${\mathcal H}$ proceeds by means of the Floquet 
approach, as successively developed in, among others, 
Refs.~\cite{AutlerTownes55,Shirley65, Zeldovich67, Zeldovich73,Sambe73,
BaroneEtAl77,FainshteinEtAl78}, presently providing the theoretical backbone 
for the discussion of Floquet time crystals~\cite{KeyserlingkEtAl16,
ElseEtAl16,RussomannoEtAl17,SachaZakrzewski18,SuraceEtAl19,KhemaniEtAl19,
GuoLiang20,ElseEtAl20,PizziEtAl21,ZalatelEtAl23}. Essentially, Floquet states 
constitute solutions of the time-dependent Schr\"odinger equation that have 
the temporal Bloch form 
\be
	|\psi_\ell(t)\rangle = |u_{\ell,0}(t)\rangle 
	\exp\left(-\ri\varepsilon_{\ell,0} t/\hbar\right)
\label{eq:FST}
\ee
with $T$-periodic Floquet function 
$|u_{\ell,0}(t)\rangle = |u_{\ell,0}(t+T)\rangle$ and 
quasienergy~$\varepsilon_{\ell,0}$. Here, we assume that the set of all these 
functions with discrete state index $\ell = 1,2,3,\ldots$ provides a complete 
orthonormal system in ${\mathcal H}$ at each instant~$t$, as is automatically 
guaranteed if  ${\mathcal H}$ is of finite dimension. In particular, this 
holds for the driven Bose-Hubbard dimer~(\ref{eq:DBH}) when working with a 
fixed number $N$ of particles, since then the dimension of its  state space 
${\mathcal H}_N$ amounts to ${\rm dim} \, {\mathcal H}_N = N+1$. 
The rationale for attaching the additional index ``$0$'' to both 
the Floquet functions and the quasienergies, but not to the Floquet 
states~(\ref{eq:FST}) themselves, will become evident below.

Inserting such a Floquet state into the time-dependent Schr\"odinger equation, 
its Floquet function is seen to obey the equation
\be
	\left( H(t) - \ri\hbar\frac{\rd}{\rd t} \right) | u_{\ell,0}(t) \rangle 
	= \varepsilon_{\ell,0} | u_{\ell,0}(t) \rangle \; .
\label{eq:TEE}
\ee
This appears to be an eigenvalue equation for the 
quasienergy~$\varepsilon_{\ell,0}$, but within standard quantum mechanics 
terminology it is not, since the temporal  derivative $-\ri\hbar\rd/\rd t$ is 
not a proper self-adjoint operator on ${\mathcal H}$. This seemingly formal
problem that actually has profound implications for physics is resolved by 
``lifting''  Eq.~(\ref{eq:TEE}) from ${\mathcal H}$ to an extended Hilbert 
space that has been introduced into physics by Sambe~\cite{Sambe73}, and 
now commonly is denoted as ${\mathcal K} = L^2[0,T] \otimes {\mathcal H}$ 
in the mathematical literature~\cite{Howland89,Joye94}. This extended space 
is comprised  {\em only\/} of square-integrable functions that {\em a priori\/} 
are $T$-periodic in time. The extension consists in the requirement that the 
time~$t$ play the role of an additional {\em coordinate\/} in ${\mathcal K}$, 
rather than a variable that parametrizes the flow as it does in 
${\mathcal H}$. Consequently, the scalar product on  ${\mathcal K}$ 
naturally involves the integration over the time coordinate. Following Sambe, 
this  scalar product is indicated by double brackets~\cite{Sambe73},
\be
	\lF u | v \rF = \frac{1}{T} \int_0^T \!\! \rd t \, \langle u(t) | v(t) \rangle \; ,
\ee
where $| u(t) \rangle$ and $| v(t) \rangle$ are $T$-periodic functions, 
not necessarily Floquet functions in the sense of Eq.~(\ref{eq:TEE}), and 
$\langle u(t) | v(t) \rangle $ are their scalar products on ${\mathcal H}$, 
evaluated at the instants~$t$. In accordance with this notation, such 
functions $| u(t) \rangle$ and $| v(t) \rangle$, when regarded as elements 
of  ${\mathcal K}$, are written as  $| u \rF$ and $| v \rF$, respectively.
 
The suggestive equation~(\ref{eq:TEE}) then translates into a  well-formulated
quantum mechanical  eigenvalue problem on ${\mathcal K}$ in the form
\be
	K | u \rF = \varepsilon | u \rF \; ,
\label{eq:WFP}	
\ee
where
\be 
	K = H(t) + p_t
\label{eq:QEO}
\ee
is the quasienergy operator acting on  ${\mathcal K}$, and 	
\be
	p_t = \frac{\hbar}{\ri} \frac{\rd}{\rd t}
\ee
actually serves as a proper momentum operator on ${\mathcal K}$ that is 
conjugate to the new coordinate~$t$, because of the periodic 
boundary condition imposed on all elements of ${\mathcal K}$.

Now, there is an important observation to be made. With $\omega = 2\pi/T$ 
and integer $m = 0, \pm1, \pm2,\ldots\/$, the functions
\be
	| u_{\ell,m}(t) \rangle = 
	| u_{\ell,0}(t)\rangle \exp(\ri m \omega t) \equiv | u_{\ell,m} \rF 
\label{eq:SOL}
\ee
likewise are $T$-periodic and solve Eq.~(\ref{eq:TEE}), or the eigenvalue 
problem~(\ref{eq:WFP}), with quasienergy  
\be
	\varepsilon_{\ell,m} = \varepsilon_{\ell,0} + m \hbar\omega \; ;
\label{eq:BZS}
\ee 		
for obvious reasons, we refer to the index~$m$ as the ``photon index.'' 
Clearly, such solutions~(\ref{eq:SOL}) with the same state index~$\ell$ and 
arbitrary photon index~$m$ produce {\em the same\/} Floquet 
state~(\ref{eq:FST}), but they constitute~{\em different\/} solutions of 
Eq.~(\ref{eq:WFP}), all of them being individually required for the 
completeness relation on ${\mathcal K}$. Hence, a quasienergy with state 
index~$\ell$ is not a mere number, but a set of equivalent representatives 
spaced by~$\hbar\omega$. The quasienergy spectrum therefore extends itself 
periodically over the whole energy axis from $-\infty$ to $+\infty$ even if 
${\mathcal H}$ is of finite dimension, with one representative of the 
quasienergy of each state located in each interval  of width $\hbar\omega$; 
with wording borrowed from solid-state physics, such an interval is termed 
Brillouin zone. The fact that the quasienergy spectrum is unbounded not only 
from above, but also from below, ties in with the observation that the 
momentum operator~$p_t$ appears only linearly in the quasienergy operator~$K$, 
reminiscent of eigenvalue equations in relativistic quantum mechanics. 
Moreover, this unboundedness implies that, strictly speaking, there is no 
Floquet ground state.

A further noteworthy feature arises for systems with infinitely many states, 
each of them placing one quasienergy representative in each Brillouin zone 
when driven periodically in time, so that the quasienergies may cover these 
zones densely. It is then a fairly difficult task to decide whether or not the 
quasienergy spectrum has an absolutely continuous component, giving rise to 
diffusive energy growth~\cite{Howland89, Joye94}. Taking up the path-directing  
mathematical analysis by Bellissard~\cite{Bellissard85}, this question has 
become known as  the quantum stability problem~\cite{Combescure88}. 
This needs to be kept in mind when considering the driven Bose-Hubbard
dimer~(\ref{eq:DBH}) in the formal limit $N \to \infty$. Some particular
ramifications of this stability problem for asymptotically large, but still
finite~$N$ will be elucidated in the Appendix.

Barring these subtleties, the underpinning for the line of reasoning adopted 
in the following stems from the observation that the quasienergy eigenvalue 
equation~(\ref{eq:WFP}) is a {\em bona fide\/} conceptual equivalent of the 
time-independent Schr\"odinger equation for energy eigenvalues and 
eigenstates. This implies that one can transfer many techniques  known from 
{\em time-independent\/}  quantum mechanical problems, such as steady-state 
perturbation theory, to {\em periodically time-dependent\/} 
systems~\cite{Sambe73}.  Thus, a general strategy for dealing with 
Floquet-type systems in an analytical  manner requires to {\em (i)\/ }first 
lift the problem of interest to the extended Hilbert space~${\mathcal K}$,  
{\em (ii)\/} apply known techniques there, and then {\em (iii)\/} project back 
to the physical  space~${\mathcal H}$.

\section{Intermediates: Semiclassical quantization of integrable Floquet 
systems}
\label{S_3}

We will now follow an analogous route --- lifting to an extended classical
phase space, applying standard quantization procedures there, and then 
projecting back to the physical space --- in order to obtain a semiclassical 
approximation to quasienergies for fully integrable periodically 
time-dependent  systems~\cite{BreuerHolthaus91}. As a reminder, 
let us recall the semiclassical Einstein-Brillouin-Keller (EBK) quantization
procedure~\cite{Gutzwiller90} of an integrable classical time-independent 
system with $f$~degrees of freedom deriving from a Hamiltonian function 
$H_{\rm cl}(p,q)$, where we write~$p$ for the momentum variables 
$p_1, \ldots, p_f$, likewise $q$ for their conjugate coordinates: 
Inte\-grability implies that the system's phase space~${\mathcal P}$ is 
completely stratified into $f$-tori $\mathbb T_f$ that are invariant under 
the Hamiltonian flow~\cite{Gutzwiller90,Arnold78,AbrahamMarsden08}. Those 
tori that can ``carry'' a quantum energy eigenstate are singled out by the 
Bohr-Sommerfeld-like conditions 
\be
	\oint_{\gamma_k} \!\! p \rd q 
	= 2\pi\hbar\!\left( n_k + \frac{{\rm ind} \,\gamma_k}{4} \right) \; , 
\label{eq:EBK}
\ee	
where $\gamma_k$ ($k = 1,\ldots,f$) denote the topologically inequivalent
contours around such a torus, $n_k$ are integer quantum numbers,  and
 ${\rm ind} \, \gamma_k$ is a Maslov index that accounts for the turning
 points of the respective contour~\cite{Gutzwiller90}. After transforming
 $H_{\rm cl}(p,q)$ to action variables, and inserting the actions of the
 tori selected by the conditions~(\ref{eq:EBK}) into the transformed function,
 one thus obtains semiclassical approximations to the energy eigenvalues of
 the classical system's quantum counterpart.

When adapting this procedure to integrable peri\-odi\-cally 
time-dependent systems governed by a Hamiltonian function 
$H_{\rm cl}(p,q,t) =H_{\rm cl}(p,q,t+T)$,
one again requires an even-dimensional phase space with pairs of canonically
conjugate momentum and position variables. Therefore, the time~$t$ that 
merely parametrizes the flow in the system's actual phase space~${\mathcal P}$ 
is being considered as a coordinate and augmented by a canonically conjugate 
momentum variable~$p_t$, in precise analogy to the viewpoint adopted in 
quantum mechanics when proceeding from Eq.~(\ref{eq:TEE}) to the eigenvalue 
equation~(\ref{eq:WFP}), altogether providing an even-dimensional extended 
phase space denoted here as ${\mathcal T} = T \otimes {\mathcal P}$. 
Consequently, the correspondent of the quasienergy operator~(\ref{eq:QEO}) 
now is the classical quasienergy function 
$K_{\rm cl} = H_{\rm cl} + p_{\rm t}$. With the original time~$t$ being a 
coordinate on equal footing with~$q$, one is forced to introduce a new  
quasitime~$\tau$ in order to parametrize the flow generated by~$K_{\rm cl}$
in~${\mathcal T}$, so that the Hamiltonian equations read 
 \begin{eqnarray}
 	\frac{\rd q}{\rd \tau} & = & 
 	\phantom{-}\frac{\partial K_{\rm cl}}{\partial p} 
 	= \phantom{-}\frac{\partial H_{\rm cl}}{\partial p}
 \nonumber \\
 	\frac{\rd p}{\rd \tau} & = & 
 	-\frac{\partial K_{\rm cl}}{\partial q} 
 	= -\frac{\partial H_{\rm cl}}{\partial q}		
 \label{eq:HS1}
 \end{eqnarray}	
for the old pairs of positions and momenta, and
\begin{eqnarray}
 	\frac{\rd t}{\rd \tau} & = & 
 	\phantom{-}\frac{\partial K_{\rm cl}}{\partial p_t} 
 	= \phantom{-}1
 \nonumber \\
 	\frac{\rd p_t}{\rd \tau} & = & 
 	-\frac{\partial K_{\rm cl}}{\partial t} 
 	= -\frac{\partial H_{\rm cl}}{\partial t}		
 \label{eq:HS2}
 \end{eqnarray}	
 for the new one. With respect to the original dynamics generated by 
 $H_{\rm cl}$ in~${\mathcal P}$, this system possesses a straightforward
 interpretation: The first of the strange-looking equations~(\ref{eq:HS2})     
 will allow one to identify the auxiliary time~$\tau$ with the actual 
 time~$t$, so that the proper Hamiltonian equations pertaining to 
 $H_{\rm cl}$ are recovered from the set~(\ref{eq:HS1}). The second of the 
 equations~(\ref{eq:HS2}) then implies that $K_{\rm cl}$ is a constant of
 motion in~${\mathcal T}$ that we designate as~$\varepsilon$,   
 \be
 	K_{\rm cl}(p,q,p_{t},t) = \varepsilon \; ,
 \label{eq:COM}
 \ee
 constituting the classical analog of the quasienergy. Indeed, the 
 conservation of $K_{\rm cl}$ is already evident from the fact that the 
 latter is autonomous,  that is, it does not depend on~$\tau$. Observe, 
 however, that the unphysical classical $\varepsilon$ can be given any value 
 by fixing  an arbitrary initial value for  $p_t$.
 
 Proposing mere integrability of this extended system does not suffice. We 
 also have to postulate periodic boundary conditions in~$t$ for the invariant 
 manifolds in~${\mathcal T}$,  again in analogy to the periodic boundary 
 conditions imposed on the elements~$|u\rF$ of the extended Hilbert 
 space~${\mathcal K}$. Here, we presuppose that the manifolds inherit the 
 period~$T$ of their Hamiltonian, and therefore identify the coordinate 
 $t= 0$ with $t = T$. Note that it is this seemingly natural assumption that 
 will  be relaxed in the following Sec.~\ref{S_4}. With this proviso, we obtain 
 $(f+1)$-tori $\mathbb T_{f+1}$ as required for EBK quantization, so that the 
 extensions of the standard conditions~(\ref{eq:EBK}) take the form   
\be
	\oint_{\gamma_k} \!\! \big(p \rd q + p_t \rd t \big)  
	= 2\pi\hbar\!\left( n_k + \frac{{\rm ind} \, \gamma_k}{4} \right)
\label{eq:EQC}
\ee	
with $k = 1, \ldots,f+1$ in order to account for the added degree of freedom.
From here we return to the physical phase space~${\mathcal P}$ with time~$t$
as a flow parameter. This means to identify~$\tau$ with~$t$, to get rid of 
$p_t$, and to cut the $(f+1)$-dimensional tori in such a way that they reduce 
to $f$-tori that flow in time,  termed $\mathbb T_f^+$. To these 
purposes we shift the contours $\gamma_k$ with $k = 1,\ldots,f$ into a 
hyperplane $t = const.$ This implies $\rd t = 0$, so that the first~$k$ 
conditions~(\ref{eq:EQC}) for the semiclassical Floquet states re-acquire the 
familiar form~(\ref{eq:EBK}). The remaining condition is brought back to 
${\mathcal P}$ by exploiting the insight that the quasienergy function 
$K_{\rm cl}$ is a constant of motion in~ ${\mathcal T}$,  as expressed by 
Eq.~(\ref{eq:COM}), giving $p_t = \varepsilon - H_{\rm cl}(p,q,t)$. We then 
denote the periodic contour~$\gamma_{f+1}$ that is led along $\mathbb T_f^+$ 
in time as~$\gamma_t$ and observe   ${\rm ind} \, \gamma_t = 0$, since there 
are no ``turning points in time.'' Renaming the corresponding integer quantum 
number~$n_{f+1}$ as~$m$, we now have
\be
	\int_{\gamma_t} \!\! \big(p \rd q -H_{\rm cl} \rd t \big) 
	+ \varepsilon T = 2\pi\hbar m \; ,
\ee	
yielding
\be
	\varepsilon = -\frac{1}{T} \! \int_{\gamma_t} \!\!
	\big( p \rd q -H_{\rm cl} \rd t \big) + m\hbar\omega 
\label{eq:AQR}
\ee
with $\omega = 2\pi/T$. This finally is the reward of the painstakingly 
tedious above reasoning: Besides the standard conditions~(\ref{eq:EBK}), 
there is the additional rule~(\ref{eq:AQR}) that provides a semiclassical 
approximation to the  quasienergies. Remarkably, this rule  already 
accounts for the familiar arrangement~(\ref{eq:BZS}) of the quasienergy 
spectrum in Brillouin zones of width $\hbar\omega$, with $m$ serving as 
photon index. These combined quantization conditions indeed furnish the 
correct quantum mechanical quasienergies of the periodically driven harmonic 
oscillator~\cite{BreuerHolthaus91}. From here on, the construction of the 
semiclassical Floquet states in a WKB-type manner parallels the construction 
of semiclassical energy eigenstates~\cite{BreuerHolthaus91,KellerRubinow60}, 
but the technical details of this procedure are not needed for our present 
purposes.

\section{Synthesis: Pre-Floquet states and dynamical tunneling}
\label{S_4}

The above semiclassical approach to quasienergies and Floquet wave functions 
will now be employed to investigate subharmonic response of the Bose-Hubbard
dimer to periodic driving. To this end, we will first consider the system's 
classical-like mean-field dynamics, and then ``requantize'' the 
almost integrable component of the latter by means of the 
relations~(\ref{eq:EBK}) and~(\ref{eq:AQR}).

Following Refs.~\cite{SmerziEtAl97,RaghavanEtAl99}, the mean-field
approximation to the system~(\ref{eq:DBH}) is obtained by replacing 
operator pro\-ducts $ \ad_i\ab_j$, when acting on an $N$-particle space 
${\mathcal H}_N$, by $N  c_i^\ast  c_j$, and decomposing the $c$-number 
amplitudes~$c_j$  into absolute values and phases according to
\be
	c_j(\tau) = | c_j (\tau) | \exp\!\big(\ri\theta_j(\tau)\big) \; .
\ee	
Defining the population imbalance
\be
	p = | c_1 |^2 - | c_2 |^2
\label{eq:PIB}	
\ee
together with the relative phase
\be
	\varphi = \theta_2 - \theta_1 \; ,
\label{eq:REP}
\ee	
the mean-field equations of motion then are equivalent to the equations of
motion furnished by the dimensionless classical single-particle Hamiltonian 
function 
\be
	H_{\rm mf}(\tau) = \alpha p^2 - \sqrt{1 - p^2}  \, \cos(\varphi)	 	
	+ 2 \frac{\mu}{\Omega}  \, p \, 
	\sin\!\left(\frac{\omega}{\Omega} \tau\!\right)  	
\label{eq:HMF}
\ee
that conforms to a periodically driven pendulum with momentum-shortened
length~\cite{SmerziEtAl97,RaghavanEtAl99}. Here, we use the time variable 
$\tau = \Omega t$, and invoke the parameter  
\be
	\alpha = N\kappa/\Omega 
\ee
that is inversely proportional to the pendulum mass. 
Multiplication of $H_{\rm mf}$ by $N\hbar\Omega$ then procures approximate
energies pertaining to the actual $N$-particle quantum system~(\ref{eq:DBH}).   
Importantly, when the mean-field dynamics are compared to those of the 
$N$-particle system for various~$N$, the ratio $\kappa/\Omega$ has to be 
adjusted such that the numerical value of $\alpha$ remains unchanged. Hence,
when the strength $\hbar\Omega$ of the tunneling contact is kept constant
while the particle number  $N$ is made large, as conducted in the present 
work, the interparticle interaction strength $\hbar\kappa$ has to be reduced 
accordingly. With this proviso, the very same phase-space structure 
can be juxta\-posed to the quantum $N$-particle dynamics for {\em all\/}~$N$, 
pictorially meaning that the phase space can be ``filled'' with an arbitrary 
number of particles.  

\begin{figure}[t]
\centering
\vspace*{2.6ex}
\includegraphics[width=1.0\linewidth]{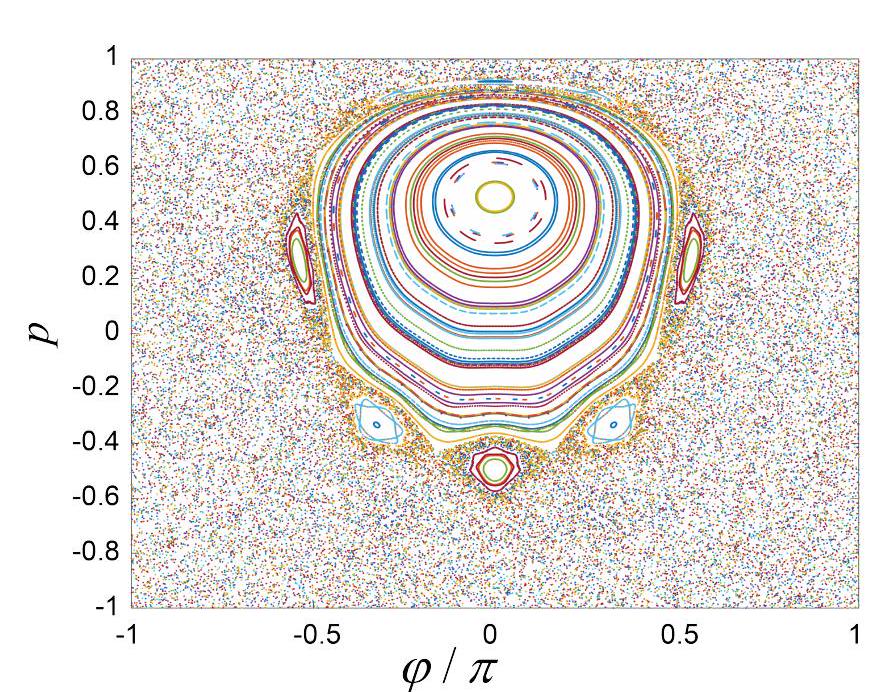}
\caption{Poincar\'e map generated at time $t = 0$ by the peri\-odically driven 
pendulum~(\ref{eq:HMF}), providing the mean-field description of the many-body
system~(\ref{eq:DBH}), with parameters $\alpha = 0.92$, $\mu/\Omega = 0.40$, 
and $\omega/\Omega = 1.90$.} 
\label{F_2}
\end{figure}

In Fig.~\ref{F_2}, we depict a Poincar\'e map for this driven 
pendulum~(\ref{eq:HMF}) that visualizes the intersection of the Hamiltonian 
flow at time $t = 0$ with the  phase-space plane~\cite{Gutzwiller90,LiLi92}. 
This map has been produced in the usual manner by integrating Hamilton's 
equations for a set of appropriately placed initial conditions over a large 
number of  driving periods~$T = 2\pi/\omega$, and recording the image points 
once per period at $t = 0 \bmod T$. In addition, we have given each sequence
of successors originating from one of the initial phase-plane points its own 
color~\cite{Data}. Para\-meters employed here are  $\alpha = 0.92$,  
$\mu/\Omega = 0.40$, and $\omega/\Omega = 1.90$, the same as 
in Fig.~\ref{F_1}. As expected, this map features the coexistence of 
regular and chaotic motion that is typical for nonlinear Hamiltonian 
systems~\cite{LiLi92}. The large island of close-to-regular motion observed 
here stems from the principal resonance that occurs when the time required for 
one oscillation of  the undriven pendulum about matches one driving period, 
so that the elliptic fixed point in its center indicates a stable $T$-periodic 
orbit. Given the pendulum's single degree of freedom, the invariant curves 
surrounding this fixed point represent $1$-tori $\mathbb T_1$ in the language 
of the preceding section; when continued in time, these flowing curves 
generate the $T$-periodic tubes~$\mathbb T_1^+$ required by the semiclassical 
rules~(\ref{eq:EBK}) and~(\ref{eq:AQR}).

We now set out to verify that these rules, mathematically designed for fully 
integrable systems, also capture the  exact $N$-particle quantum dynamics 
under the pseudo-integrable conditions prevailing in this island. Expressed in 
terms of the variables $p,\varphi,\tau$ appearing in the dimensionless 
Hamiltonian function~(\ref{eq:HMF}), they take the forms
\begin{eqnarray}
	\oint_{\gamma_1} \!\!  p \rd\varphi & = &
	2\pi\hbar_{\rm eff} \! \left( n + \frac{1}{2} \right) \; ,
\nonumber \\
	\frac{\varepsilon}{N\hbar\Omega} & = &
	-\frac{1}{\Delta\tau} \! \int_{\gamma_\tau} \!\!
	\big( p \rd \varphi  - H_{\rm mf} \rd \tau \big) 
	+ m\hbar_{\rm eff}\frac{2\pi}{\Delta\tau} \; . \phantom{xx}
\label{eq:SCS}
\end{eqnarray}
The effective Planck constant $\hbar_{\rm eff}$ introduced here 
is determined by the requirement that the total area 
$\Delta p \times \Delta \varphi$ of the phase-space plane has to accommodate 
the $N+1$ Floquet states of the quantum system~(\ref{eq:DBH}), each of these
occupying an area $2\pi\hbar_{\rm eff}$, thus demanding 
$2 \times 2\pi = 2\pi\hbar_{\rm eff}(N+1)$, or   
\be
	\hbar_{\rm eff} = \frac{2}{N}
\label{eq:EPC}
\ee
for largish~$N$. Moreover, we have inserted  ${\rm ind} \, \gamma_1 = 2$ 
for the two turning points  of  a contour~$\gamma_1$ around $\mathbb T_1^+$,
 and have written $\Delta\tau = 2\pi\Omega/\omega$ for the scaled cycle  
 duration.

Next, we utilize the coherent $N$-particle states~\cite{Radcliffe71}
\be
	|\vartheta,\varphi\rangle_{\!N} = \frac{1}{\sqrt{N!}} 
	\left(A^\dagger(\vartheta,\varphi)\right)^{\!N} 
	| \rm{vac} \rangle \; ,
\label{eq:CNP}
\ee
where the creation operators 
\be 
	A^\dagger(\vartheta,\varphi) = \cos\!\frac{\vartheta}{2} \,\ad_1 
	+ \sin\!\frac{\vartheta}{2} \, \re^{\ri\varphi} \, \ad_2
\ee
act on the empty-dimer state $| \rm{vac} \rangle$, so that the specific 
population imbalance~(\ref{eq:PIB}) of such a state is given by
$p = \cos^2(\vartheta/2) - \sin^2(\vartheta/2) = \cos\vartheta$,
 while $\varphi$ coincides with the relative phase~(\ref{eq:REP}).
 Hence, the squared scalar product
\be
	Q^{(N)}_{|\psi\rangle}(p,\varphi) = 
	\big| \langle \psi | \vartheta,\varphi \rangle_{\!N} \big|^2  
\label{eq:HPN}
\ee
reveals how strongly a given $N$-particle state $|\psi\rangle$ is associated 
with the phase-space point $(p = \cos\vartheta,\varphi)$; computation of this 
quantity~(\ref{eq:HPN}) for all $-1 \le p \le +1$ and  $-\pi \le \varphi \le +\pi$ 
provides a  Husimi projection of that quantum state onto the classical 
phase-space plane.
 
\begin{figure}[t]
\centering
\includegraphics[width=1.0\linewidth]{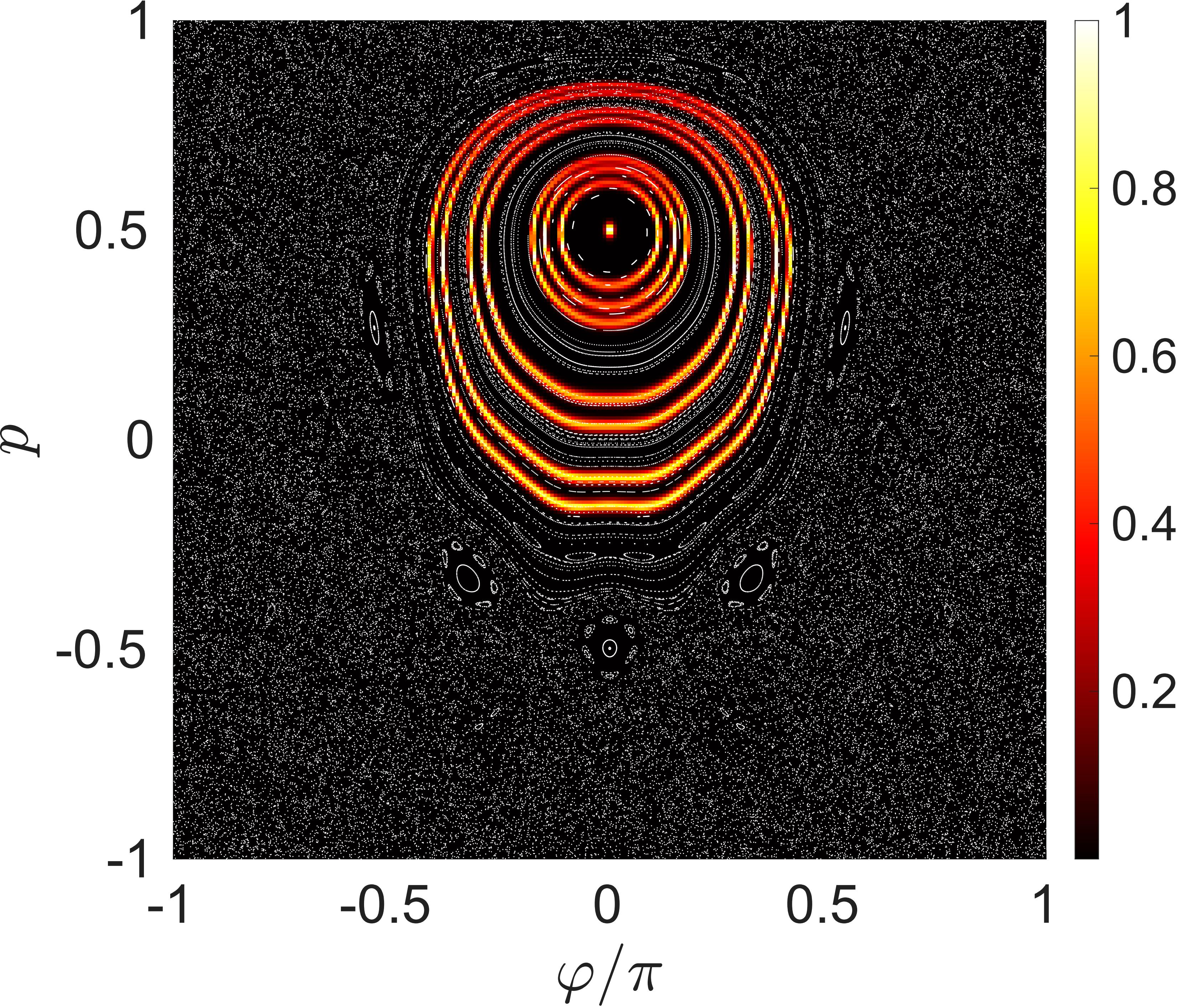}
\caption{Color-coded Husimi projections~(\ref{eq:HPN}) of eight Floquet states 
$|\psi\rangle = |u(0)\rangle$ for $N = 10\,000$ particles, intersected at time 
$t=0$, onto the surface of section shown in Fig.~\ref{F_2}, here depicted in
black and white. Observe that each of these states is localized on 
its respective closed contour $\gamma_1$ selected by the upper of the 
conditions~(\ref{eq:SCS}). Their semiclassical quantum numbers are $n = 0$,  
$109$, $193$, $275$, $767$, $971$, $1414$, and $1672$  (inner to outer).} 
\label{F_3}
\end{figure}

In Fig.~\ref{F_3}, we display such color-coded Husimi 
projections~(\ref{eq:HPN}) of eight numerically computed Floquet states 
$|\psi\rangle = |u(0)\rangle$ for $N = 10\,000$ particles. One 
of these states emerges as the light spot in the middle of the central island, 
right around the elliptic fixed point, each of the others as one of the 
surrounding  light closed curves covering their respective contour $\gamma_1$ 
as selected by the first of the conditions~(\ref{eq:SCS}). These states are 
superimposed on the mean-field Poincar\'e section, indicated here in black 
and white only. While Floquet states normally cannot be ordered with respect 
to the magnitude of  their quasienergies, because of  the unbounded 
Brillouin-zone-like quasienergy spectrum, Floquet states that are 
semiclassically associated with an island of almost regular mean-field motion 
can be well ordered with respect to their semiclassical  quantum 
numbers~\cite{SeligmannHolthaus25}. Referring to the quantum number $n = n_1$ 
employed in the upper of the scaled conditions~(\ref{eq:SCS}), the states 
portrayed in Fig.~\ref{F_3} carry the labels $n = 0$, $109$, $193$, $275$, 
$767$, $971$, $1414$, and $1672$, respectively (inner to outer). Hence, the 
Floquet state $n = 0$ that adheres most closely to the elliptic periodic orbit 
constitutes the resonance-induced effective  ground state of the main regular 
island. The excited states, viewed here at $t = 0$ only, likewise cling to 
their respective invariant circles~$\mathbb T_1$; when continued in time, they 
stick to the emanating tubes $\mathbb T_1^+$. Thus, Fig.~\ref{F_3} provides a 
visible testimony of  the fact that exact Floquet states that occupy 
predominantly regular regions of phase space are attached to their mean-field 
tubes in a semiclassical manner. Having descended before from the quantum 
mechanical $N$-particle level to a classical-like mean-field description, we refer 
to the return from that description to the full $N$-particle dynamics by means 
of the conditions~(\ref{eq:EBK}) and~(\ref{eq:AQR}) as 
requantization~\cite{SeligmannHolthaus25}.

Still, from the viewpoint of rigorous mathematics, there is an
objection that should be taken into account at this place. If the mean-field 
motion within the almost regular islands indeed were fully integrable,  the 
Bohr-Sommerfeld-like rules would always find their exact contours~$\gamma_1$. 
Actually, however, there is fine-scale chaos even within the  seemingly 
regular islands, due to the destruction of all those tubes for which the ratio 
between the driving frequency and the unperturbed pendulum  frequency is 
rational, giving way to chains of alternating elliptic and hyperbolic fixed 
points with the associated homoclinic tangles, in accordance with the 
Poincar\'e-Birkhoff theorem~\cite{Gutzwiller90,AbrahamMarsden08,LiLi92}.
 Thus, there are  chaotic dynamics ``squeezed'' between the remaining 
 preserved tubes, self-similar on all  (infinitesimally fine) scales, although 
 these are not accessible to finite-precision arithmetics. Hence, those tubes 
 which fulfill the quantization condition identically (in the absolute 
 mathematical sense) may not exist, even if there are others that come 
 arbitrarily close. As long as $N$ remains finite, so that each quantum state 
 occupies a finite area $2\pi\hbar_{\rm eff}$, the $N$-particle system will be 
 unable to resolve such hyperfine phase-space structures,  which therefore 
 remain practically undetectable. On the other hand, this implies that the 
 formal limit $N \to \infty$ is not well controlled. 
 
 For driven nonlinear oscillators with only one degree of freedom, such as 
 those considered here, the preserved tubes represent impenetrable barriers 
 for each trajectory, because of the uniqueness theorem for ordinary 
 differential equations. Therefore, a mean-field trajectory starting within 
 a mainly regular island, chaotic or not, remains within this island  forever. 
 This is different for quantum mechanical wave packets, as discussed in the 
 Appendix. 
  
\begin{figure}[b]
\centering
\includegraphics[width=1.0\linewidth]{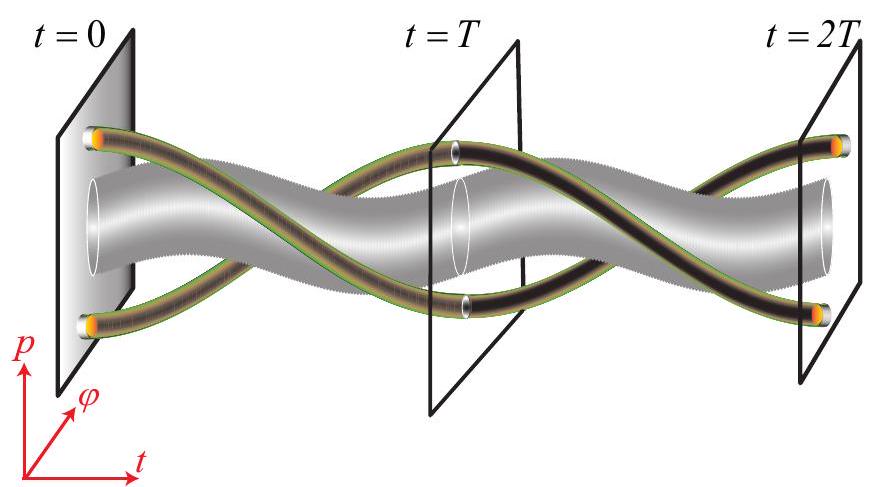}
\caption{Phase-space geometry pertaining to a hypothetical $1:2$ resonance
(schematically). The central $T$-periodic mean-field tube provides a proper 
approximate $N$-particle Floquet state upon semiclassical ``requantization.''
The two $2T$-periodic tubes winding around it yield two $2T$-periodic 
pre-Floquet states. Taking even and odd superpositions of these, thus
accounting for dynamical tunneling between them, gives two approximate
Floquet states.\/}    
\label{F_4}
\end{figure}

Around the main regular island, Fig.~\ref{F_2} also reveals a chain of six 
smaller secondary islands, one of them squeezed at the top. The alternating
coloring within this chain indicates that it disaggregates into two separate
subchains, each consisting of three islands, so that the elliptic fixed point 
in the center of each island stems from a periodic orbit that closes in 
itself after three driving periods. Therefore, the invariant curves 
surrounding them represent sections of two disconnected sets of $3T$-periodic  
tubes with the Poincar\'e plane. To make the underlying geometry more clear, 
we sketch in Fig.~\ref{F_4} the hypothetical case of a single $1:2$ resonance. 
The large central tube depicted there visualizes a $T$-periodic flowing contour 
$\mathbb T_1^+$ as considered before. Winding around it there are two narrower 
tubes that are displaced against each other by  one driving period~$T$, 
otherwise being identical, and therefore provide two intersections with the 
Poincar\'e plane. The simultaneous existence of tubes with different periods 
is possible in nonintegrable systems only, since such tubes necessarily 
have to be separated by zones of chaotic motion within which no invariant 
manifolds exist. Disregarding these zones, one can now apply the semiclassical 
rules~(\ref{eq:SCS}) to each of these secondary tubes individually, 
in the nonrigorous sense discussed above, replacing the scaled 
driving period $\Delta\tau$ by the tube period $2\Delta \tau$. As required by 
the second of these rules, the width of the entailing quasienergy Brillouin 
zone then shrinks from $\hbar\omega$ to $\hbar\omega/2$.  Moreover, 
since there are two equivalent tubes, the semiclassical states obtained 
by single-tube quantization appear in doublets with still identical 
quasienergies. But evidently each of these $2T$-periodic semiclassical states 
alone cannot yet approximate a proper quantum mechanical Floquet state, since 
Floquet functions inevitably are strictly $T$-periodic. For this reason, we 
denominate states constructed by semiclassical quantization of single tubes 
possessing a period other than~$T$ pre-Floquet states.

The situation encountered here closely parallels the double-well paradigm 
alluded to in the Introduction. Semiclassical quantization of the motion in 
each well of a symmetric double-well potential, disregarding tunneling through 
the barrier, provides pre-eigenstates with identical energies. However, these
states do not respect the actual reflection symmetry of the system. This 
symmetry is restored when accounting for quantum tunneling through the 
barrier by taking even and odd superpositions of the pre-eigenstates, 
thereby introducing a tiny energy splitting between the members of each 
doublet~\cite{LaLiQM}. By ana\-logy, even or odd superpositions of  the 
pre-Floquet states derived from a $1:2$ resonance acquire the proper 
translational symmetry in time. Here, we do not find tunneling through a 
barrier, but dynamical tunneling through a chaotic zone of phase space between 
symmetry-related islands, akin to the quantum dynamical tunneling in bound 
states pioneered by Davis and Heller~\cite{DavisHeller81}. By the same token, 
such even or odd super\-positions of pre-Floquet states also give rise to a 
tiny  quasienergy splitting. This implies that initial states $|\psi(0)\rangle$ 
prepared in a single pre-Floquet state will tunnel from one tube to the other 
on a rather long timescale that is inversely proportional to that splitting. 
For the paradigmatic case of a $1:2$ resonance, the splitting of 
each individual Floquet-state doublet can approximately be expressed analytically in 
terms of  characteristic values of the Mathieu equation~\cite{Holthaus95,
HolthausFlatte94,FlatteHolthaus96}; higher resonances are treated 
accordingly. In particular, dynamical tunneling within the driven Bose-Hubbard
dimer~(\ref{eq:DBH}) has been studied in some detail both analytically and
numerically for moderate particle numbers~\cite{GertjerenkenHolthaus14}.

\begin{figure}[t]
\centering
\includegraphics[width=0.7\linewidth]{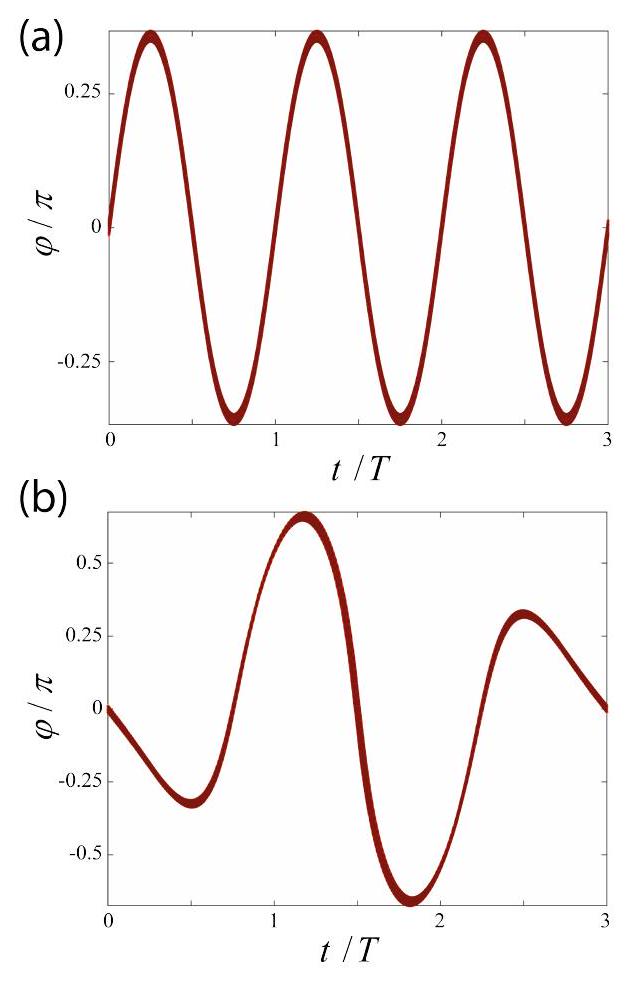}
\caption{(a) Projection of a tube obtained by following an invariant contour 
surrounding the central elliptic fixed point of the main regular island 
depicted in Fig.~\ref{F_2} in time. Such tubes are $T$-periodic, providing 
$T$-periodic Floquet states upon semiclassical quantization.
(b) Projection of a tube generated by following a contour surrounding 
the central elliptic fixed point of the lowest secondary island observed 
in Fig.~\ref{F_2} in time. Such tubes are $3T$-periodic, and therefore provide
$3T$-periodic pre-Floquet states that effectuate the $1:3$ subharmonic 
clocking recognized in Fig.~\ref{F_1}.}  
\label{F_5}
\end{figure}

Coming back to our guiding numerical example, we display in  Fig.~\ref{F_5}(a) 
the projection from $(p,\varphi,t)$ space to the $(\varphi,t)$ plane of a tube 
that emanates from a contour encircling the central elliptic fixed point 
inside the main regular island observed in  Fig.~\ref{F_2}. Such tubes are 
$T$-periodic and thus provide the scaffolds for semiclassical Floquet states 
that effectuate standard $1:1$ clocking, akin to the  wide $T$-periodic tube
sketched in Fig.~\ref{F_4}. In contrast, Fig.~\ref{F_5}(b) depicts the  
projection of a tube generated by a contour surrounding  the central 
elliptic fixed point in the lowest island of the secondary chain. Evidently, 
this projection closes in itself after three driving periods, confirming that 
the ostensive chain of six islands  actually consists of two disconnected 
three-island subchains. Thus, there are two differences in comparison to 
the pedagogical Fig.~\ref{F_4}: The tubes showing up here possess the 
period~$3T$, and there exist even two sets of such tubes. Each set thus 
provides $3T$-periodic tunneling-coupled pre-Floquet states upon
semiclassical quantization, linear superpositions of  which yield proper 
Floquet states.

\begin{figure}[t]
\centering
\includegraphics[width=1.0\linewidth]{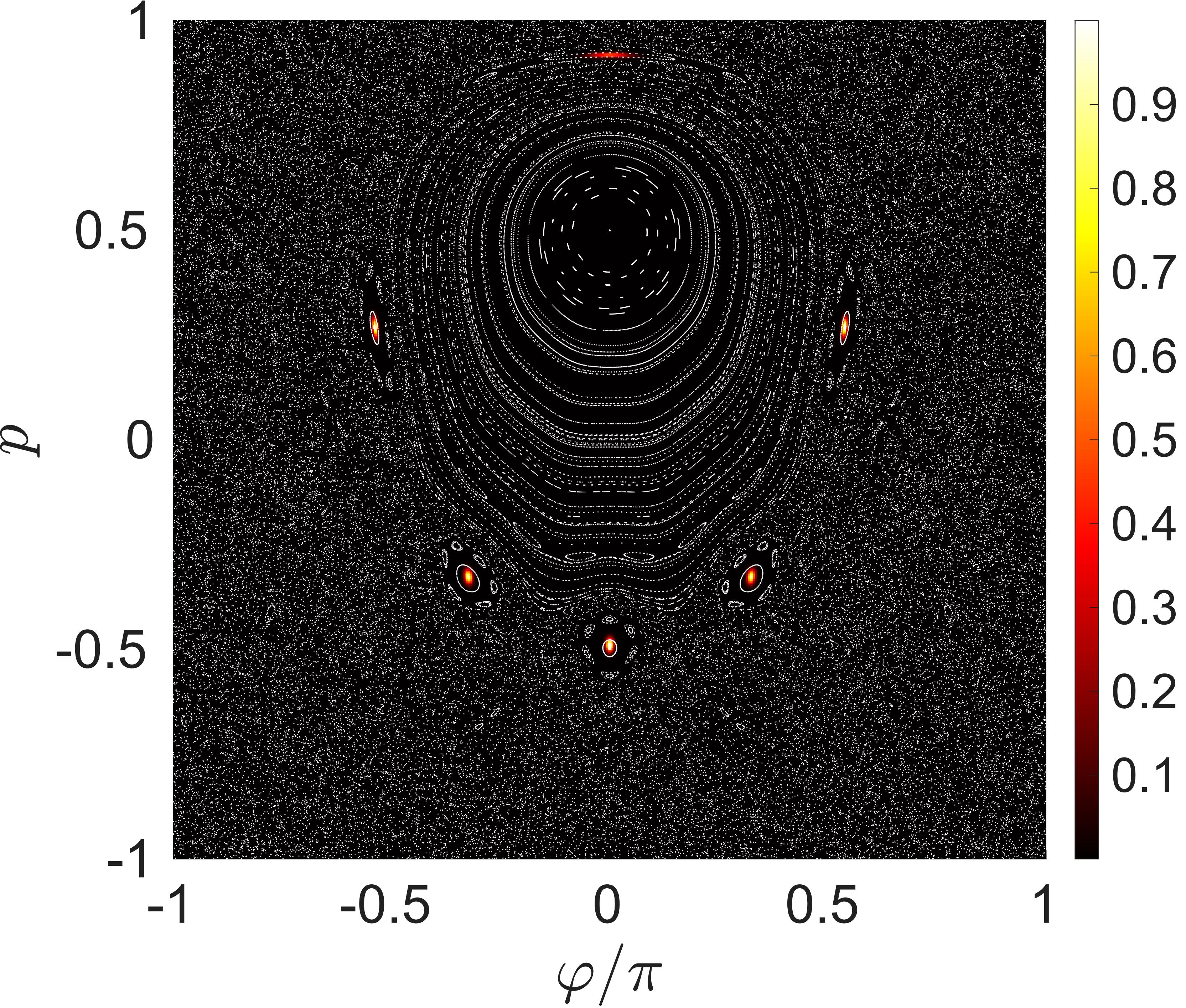}
\caption{Husimi projection of one representative of the six Floquet states 
associated with the innermost quantized tubes surrounding the elliptic 
periodic orbits belonging to the secondary six-island chain observed in 
Fig.~\ref{F_2}. The occupation of all six islands indicates not only tunnel 
coupling of three $3T$-periodic pre-Floquet states provided by one set of 
$3T$-periodic tubes,  but also hybridization with those obtained from the 
second set. The particle number is $N = 10\,000$, as in Fig.~\ref{F_3}.}  
\label{F_6}
\end{figure}

Inspecting the Husimi projection of one of the six numerically computed 
Floquet states with $N = 10\,000$ Bose particles and semiclassical  quantum 
number $n=0$ referring to the six-island chain seen in Fig.~\ref{F_2}, we find 
in Fig.~\ref{F_6}  occupation of not only one of the two disconnected subsets 
of three islands, but also of the other one. Therefore, these Floquet states 
do result not only from tunnel coupling of  three  $3T$-periodic pre-Floquet 
states, but also from hybridization with the other three. 

\begin{figure}[t]
\centering
\includegraphics[width=1.0\linewidth]{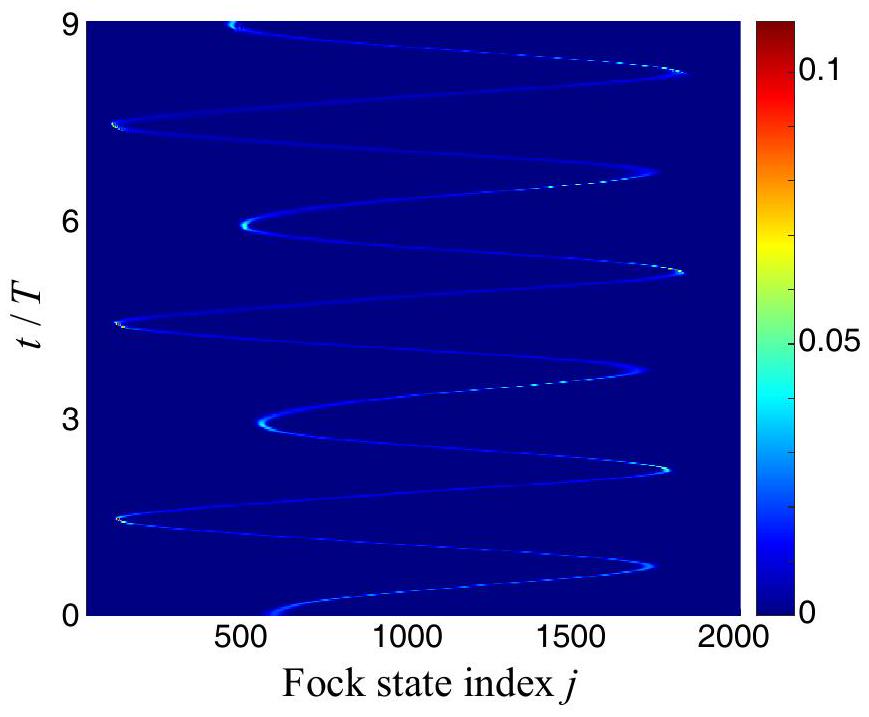}
\caption{Time evolution~(\ref{eq:TEV}) of an initial coherent 
state~(\ref{eq:CNP}) with $N = 2000$ particles and parameters 
$p = \cos\vartheta =-0.497$ and $\varphi = 0.0$ that specify the central 
elliptic fixed point in the lowest of the six secondary islands seen in 
Fig.~\ref{F_2}, viewed in the Fock basis $|j,N-j\rangle$ of the Bose-Hubbard 
dimer. This initial state $|\psi(0)\rangle$ is composed mainly of the 
$3T$-periodic pre-Floquet states associated with this island,  giving rise 
to longtime coherent subharmonic motion, and causing the $1:3$ subharmonic 
return probability $P_r(t)$ shown in Fig.~\ref{F_1}.}
\label{F_7}
\end{figure}

Notwithstanding this additional subtlety, an initial state $|\psi(0)\rangle$ 
placed on only one of the six islands is composed mainly of the associated 
$3T$-periodic pre-Floquet states, instead of $T$-periodic Floquet states, and 
therefore will feature $3T$-periodic time evolution on time scales which are 
short  in comparison with the tunneling times. This is confirmed in 
Fig.~\ref{F_7} that depicts the time evolution of an initial coherent 
state~(\ref{eq:CNP}) with $N = 2000$ particles placed right upon the elliptic 
fixed point of the lowest secondary island in Fig.~\ref{F_2}, that is, with
para\-meters $p = \cos\vartheta =-0.497$ and $\varphi = 0.0$. Here, we employ
the Fock states $|j,N-j \rangle$ of the Bose-Hubbard dimer with $j$~particles
occupying site~$1$ and, accordingly, $N-j$~particles occupying site~$2$,
and plot the color-coded evolution of their occupation probabilities
\be 
	F(j;t) = \big| \langle \psi(t) | j,N-j \rangle \big|^2 \; .
\label{eq:TEV}
\ee	 
Note that this Fig.~\ref{F_7} has been obtained by sheer numerical 
computation, not taking recourse to any semiclassical approximation or 
reasoning. The evolving state remains coherent on the short time interval 
displayed here, and exhibits almost perfect subharmonic $1:3$ motion; 
this is precisely the setting that leads to the $1:3$ clocking found in 
Fig.~\ref{F_1}.

\section{Outlook: Subharmonic response of Floquet condensates}
\label{S_5}

The pre-Floquet states expounded on in the present work are a semiclassical, 
and hence approximate concept, glossing over imperfect integrability within 
the mean-field resonance zones, and ignoring complications due to 
proliferating near-degeneracies of quasienergies. As laid out in the Appendix, 
the latter imply intricate hybridization of regular and chaotic states. 
Nonetheless, such pre-Floquet states are well suited to approach longtime, 
albeit not perpetual subharmonic response of  $T$-periodically driven 
many-boson systems on an intuitive level, giving access to wave packets that 
follow periodic mean-field orbits having a period that is an integer or 
fractional multiple of~$T$, to multiplets of quasienergies together with 
dynamical tunneling effects lifting their degeneracy, and to the ultimate 
decay of subharmonic motion. Being semiclassically constructed in a WKB-type 
way from only one of the $nT$-periodic invariant mean-field tubes pertaining 
to an island chain with $n$ individual islands, $nT$-periodic pre-Floquet 
states are related to genuine Floquet states in a manner similar to the 
relation of site-localized Wannier states to lattice-extended Bloch waves in 
solid-state physics, although the number of ``lattice'' sites is relatively 
small here, being given by the number of islands in the chain considered. The 
reasoning put forward in the Appendix suggests that pre-Floquet states span 
well-isolated subspaces of regular subharmonic motion that are  only very 
weakly coupled to a ``bath'' of chaotic states.    
 
While the mechanism for subharmonic generation investigated in the present 
case study does not involve many-body localization that would be a 
characteristic prerequisite for genuine discrete time 
crystals~\cite{ZhangEtAl17,ChoiEtAl17,KeyserlingkEtAl16,ElseEtAl16,
RussomannoEtAl17,SachaZakrzewski18,SuraceEtAl19,KhemaniEtAl19,
GuoLiang20,ElseEtAl20,PizziEtAl21,ZalatelEtAl23}, it does heavily rely on 
coherence. The return from the mean-field level to the full many-body dynamics 
by semiclassical requantization with the help of the conditions~(\ref{eq:EBK}) 
and~(\ref{eq:AQR})  can be meaningfully made only if the solutions to the 
mean-field equations of motion, effectively describing single-particle 
dynamics, represent macroscopically occupied single-particle Floquet states, 
{\em i.e.\/}, Floquet condensates. Seen from this perspective, the scenario 
exemplified in the preceding section constitutes a straightforward adaptation 
of another prescription for subharmonic generation in single-particle quantum
systems~\cite{HolthausFlatte94,FlatteHolthaus96}. A distinct new twist 
coming into play here is the appearance of an effective many-body Planck 
constant~(\ref{eq:EPC}) that is inversely proportional to the particle 
number~$N$: The larger~$N$, the smaller $\hbar_{\rm eff}$, and the finer the 
details of the mean-field phase space that the quantum $N$-particle system 
is able to resolve~\cite{SeligmannHolthaus25}. Therefore, it is the magnitude 
of~$N$ that decides whether or not a requantized mean-field tube fits 
into an island of regular motion in accordance with the first of the 
conditions~(\ref{eq:SCS}), providing a semi\-classical pre-Floquet state. 
We surmise that this feature remains decisive, with appropriate changes and 
extensions, also for experimental setups that are more complex than the 
driven Bose-Hubbard dimer~(\ref{eq:DBH}). This is of interest insofar as the 
classical Hamiltonian dynamics of  nonintegrable systems are self-similar on 
all scales~\cite{AbrahamMarsden08}. With regard to our sketchy Fig.~\ref{F_4}, 
this means that there actually exists an infinite hierarchy of ``tubes winding 
around tubes that wind around tubes.'' These should be detectable in principle 
in experiments with Floquet condensates with gradually increased numbers of 
particles and Feshbach-reduced interparticle interaction strengths, varied 
such that the product of both quantities remains constant in order to approach 
the mean-field regime. It should be kept in mind, however, that in 
contrast to time-independent Bose-Einstein condensates, which occupy true 
ground states, Floquet condensates are metastable~\cite{SeligmannHolthaus25},
as outlined in the Appendix, with lifetimes that still need to be determined.     

\begin{figure}[t]
\centering
\includegraphics[width=1.0\linewidth]{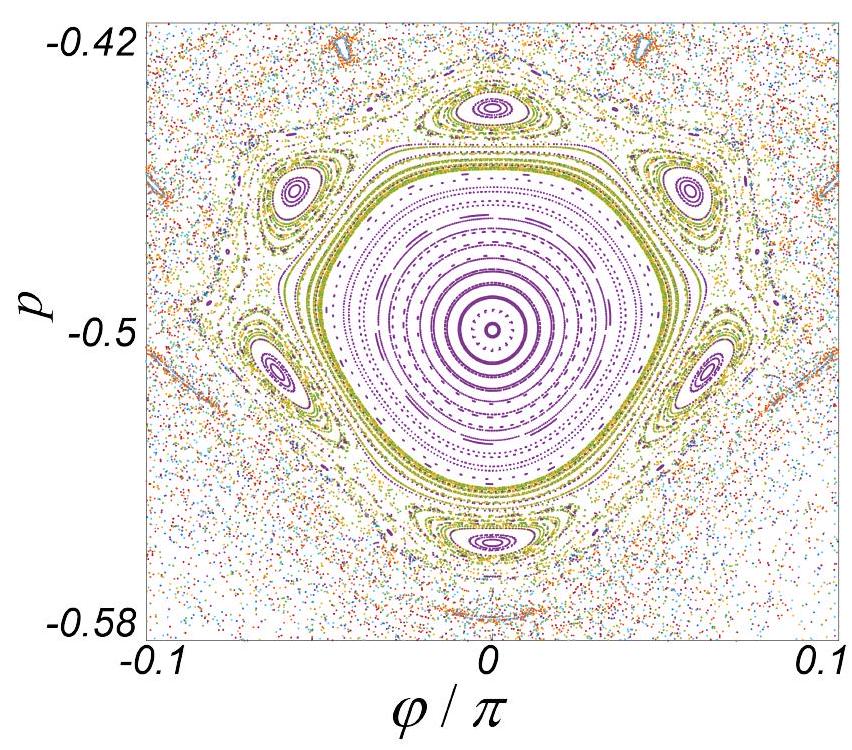}
\caption{Magnification of the lowest secondary resonant island observed
in Fig.~\ref{F_2}, revealing a surrounding further, third-order chain of six 
equivalent islands of regular motion. When $\hbar_{\rm eff}$ is made 
sufficiently small, each of these islands can host pre-Floquet states with 
period~$18T$, giving rise to $1:18$ subharmonic clocking.}
\label{F_8}
\end{figure}

\begin{figure}[t]
\centering
\includegraphics[width=1.0\linewidth]{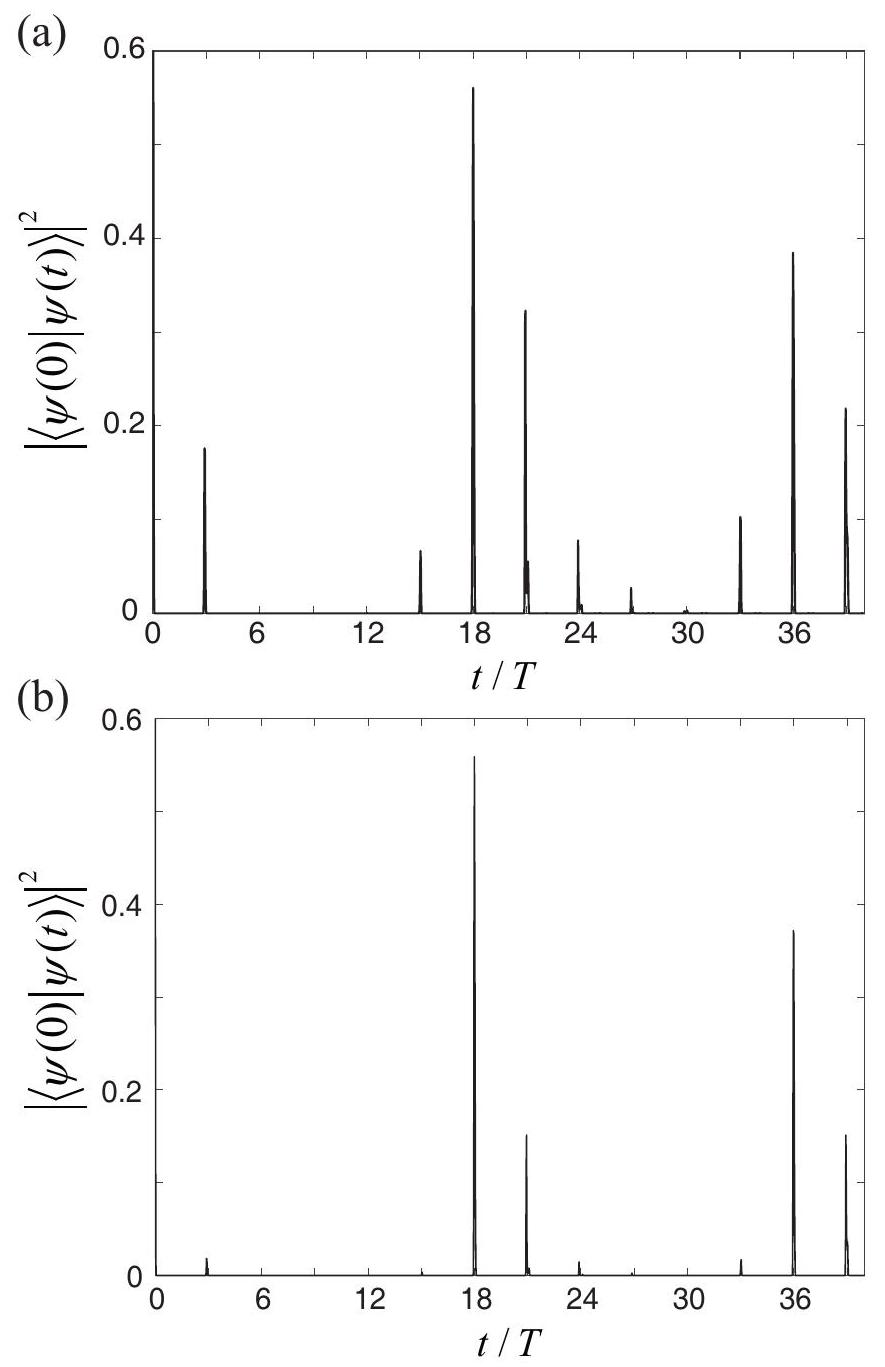}
\caption{(a) Return probability 
$P_r(t) = \left| \langle \psi(0) | \psi(t) \right|^2$
for an initial coherent state~(\ref{eq:CNP}) with $N = 2000$ particles placed 
on the lowest third-order island visible in Fig.~\ref{F_8} with $p = -0.4278$ 
and $\varphi = 0.0$. Although $\hbar_{\rm eff} = 2/N$ is not small enough to 
accommodate a quantized tube conforming to Eq.~(\ref{eq:SCS}) in that island, 
indications of the mean-field $1:18$ clocking are already showing up here.
(b) As above, but with $N = 5000$ particles. With $\hbar_{\rm eff}$ being
reduced, the sidepeaks still present in panel~(a) are substantially suppressed, 
and $1:18$ clocking unmasks itself.} 
\label{F_9}
\end{figure}

As an extension along these lines of our numerical example, we depict in 
Fig.~\ref{F_8} a magnification of one of the second-order islands previously 
observed in Fig.~\ref{F_2}: Here, one detects a surrounding third-order chain 
consisting of six islands that again produces sets of  invariant tubes; when 
$\hbar_{\rm eff}$ is made sufficiently small, these tubes in their turn host 
pre-Floquet states with return period $6 \times 3T = 18T$. Even when $N$ is 
still not large enough, so that the mean-field regime is not yet reached to the 
extent that a quantized tube would fit fully into a regular island that small, 
signs of the perfect mean-field high-order subharmonic motion can 
already mani\-fest themselves in the exact time evolution of the $N$-particle 
system. This is demonstrated in Fig.~\ref{F_9}(a) that highlights the 
return probability $P_r(t)$ for an initial coherent state~(\ref{eq:CNP}) with 
$N = 2000$ particles starting from the third-order island  with $p = -0.4278$ 
and $\varphi = 0.0$. Although the strict mean-field $1:18$ subharmonic 
clocking cannot be realized perfectly under these conditions, and signals 
related to the $3T$-periodic parent tubes still appear at most integer 
multiples of $3T$, indications of that expected high-order clocking are 
evident. Even more striking, when the particle number is increased to 
$N = 5000$, so that $\hbar_{\rm eff}$ is reduced by a factor of $0.4$, the 
side peaks are suppressed markedly and the $1:18$ clocking stands out in an 
impressive manner, as witnessed  by Fig.~\ref{F_9}(b). This example 
illustrates how to engineer high-order subharmonic response in a deliberate
and systematic way.   

Naturally, the question emerges whether concepts derived from the 
investigation of the two-site Bose-Hubbard dimer can be carried over to 
periodically driven many-body models with more sites. This question leads to 
a further demanding challenge, since even only three interacting sites without 
periodic forcing correspond to two coupled mean-field pendula endowed with a 
four-dimensional phase space~${\mathcal P}$, and therefore yield the familiar 
mixture of regular and chaotic motion by themselves~\cite{LiLi92}. Installing 
time-periodic driving then adds a new feature: Namely, while a $2$-torus 
embedded in a four-dimensional phase space divides that space into an interior 
and an exterior region, this is no longer the case in higher dimensions, 
{\em i.e.\/}, for systems with more than two degrees of  freedom. Therefore, 
such high-dimensional  tori do no longer constitute ``impenetrable diffusion 
barriers'' for chaotic motion in their extended phase space~${\mathcal T}$,
but allow for  Arnold diffusion~\cite{Arnold64,Chirikov79,BoretzReichl16}. 
Thus, a driven Bose-Hubbard trimer appears to be a prime candidate for 
studying the implications of mean-field Arnold diffusion for exact 
$N$-particle dynamics. 

Other than that, our technique indeed does apply to driven many-site systems: 
Find periodic mean-field orbits with a period other than~$T$ and check that 
these are surrounded by zones of near-integrable motion. Provided the particle 
number $N$ is sufficiently large so that ``quantizable'' tubes fit into these
zones, there will be quantum $N$-particle states that exhibit almost perfect 
subharmonic motion for many driving periods, under conditions such that exact 
many-body calculations are way beyond the capabilities of present computers.

We conclude that the experimental observation of high-order subharmonic 
motion, and of signatures of dynamical tunneling between the pre-Floquet 
states involved, would constitute a novel route towards unraveling the 
classical-quantum correspondence, as well as its limitations. Arguably, a 
major challenge for future laboratory experiments with Floquet condensates
aiming in this direction would be the preparation of the required initial
pre-Floquet states~$|\psi(0)\rangle$. This demand might potentially be matched 
by turning on the periodic drive in an adiabatic manner, possibly involving 
simultaneous variation of more than one parameter in order to guide an initial 
time-independent Bose-Einstein condensate coherently into targeted pre-Floquet 
states, leaving ample opportunities to break  genuinely new ground.

\begin{acknowledgments}
This work was supported by the Deutsche Forschungsgemeinschaft 
(DFG, German Research Foundation) through Project No.~355031190. 
We thank the members of the research group FOR 2692 for fruitful discussions. 
\end{acknowledgments}

\begin{appendix}

\section{Phase-space filling and beyond}
\label{S_A}

As discussed in Sec.~\ref{S_4}, a mean-field trajectory of the driven 
Bose-Hubbard dimer that starts in one of the almost regular islands remains 
confined to its island chain forever. In this Appendix, we argue that this is 
not the case for the exact quantum mechanical $N$-particle system, for which 
Floquet states and pre-Floquet states semiclassically connected to an island 
actually remain weakly coupled to the ``bath'' provided by the surrounding 
stochastic sea, eventually decohering harmonic and subharmonic response 
likewise. Especially this occurs for arbitrarily large, finite $N$, 
that is, when the classical mean-field phase space is filled with an 
arbitrarily large number of particles.   

 The quasienergy operator~(\ref{eq:QEO}) of the driven Bose-Hubbard dimer 
 possesses a space-time symmetry: Swapping the site indices in combination 
 with a shift in time by half a period according to
\be
	P: \left\{ \begin{array}{rcl}
	1 & \longleftrightarrow & 2	\\
	t & \longrightarrow & t + \pi/\omega
	\end{array} \right.
\label{eq:GPA}	
\ee
leaves $K$ unchanged. Clearly, this generalized parity operation~$P$ refers 
to Sambe's extended Hilbert space ${\mathcal K}$ in which $t$ is a cyclic 
coordinate,  so that  $P^2 = 1$, but not to the physical space ${\mathcal H}$. 
Hence, the eigensolutions $|u\rF$ of $K$ are even or odd under $P$. Notably,  
solutions $|u_{j,m}\rF$ that provide the same Floquet state~$j$  while  
differing in photon index~$m$ by an odd integer have different generalized 
parity, but this does not show up in the Floquet states themselves. Now the 
Neumann-Wigner noncrossing rule states that eigenvalues of a Hermitian 
operator which belong to the same symmetry class generally do not cross when 
only one system parameter is varied~\cite{NeumannWigner29}, with the possible 
exception of certain dia\-bolical points~\cite{BerryWilkinson84}.
This applies, in particular, to the eigenvalues $\varepsilon_{j,m}$ of $K$:
Upon variation of, say, the scaled driving amplitude $\mu/\Omega$, quasienergy 
representatives $\varepsilon_{j,m}$ provided by eigensolutions $|u_{j,m}\rF$ 
possessing the same generalized parity under the operation~(\ref{eq:GPA}) 
generically avoid each other, no matter whether their Floquet states are 
linked to a regular island, or to the stochastic sea. This esta\-blished fact 
has to be viewed against the finding that quasienergy eigenvalues tend to fill 
the Brillouin zones densely when $N$ is increased without bound, so that an 
arbitrarily small quasienergy interval around each representative may contain 
an arbitrarily large number of  other representatives belonging to the same 
parity class. This necessarily leads to a complex net of tiny avoided crossings 
when $\mu/\Omega$ is varied, presumably such that fine anticrossings are  
interspersed with even finer ones, and so on. In any case, the approach to 
the limit $N \to \infty$ is highly nontrivial~\cite{HoneEtAl97}. Such manifold 
avoided crossings, in their turn, inevitably imply intricate hybridization of 
the participating states, not discriminating between ``regular'' and 
``chaotic'' ones. Therefore, a wave packet initially placed on an almost 
regular island, being composed of large-$N$ Floquet states associated with 
both that island and the stochastic sea, will eventually leak out of the 
island, on time scales determined by the width of the hyperfine anticrossings.
In this regard, the quantum dynamics remain fundamentally different from their 
mean-field description even when the phase space is filled with an arbitrarily 
large number~$N$ of Bose particles.  

While this imaginary construction still does not cover the virtual limit 
``$N = \infty$'' itself that is supposed to be singular, it seems to indicate a 
precursor of the quantum stability problem that manifests itself here already 
for macroscopically large, but finite~$N$. The latter problem is driven by the 
mathematical question whether a dense point spectrum turns into a continuous 
one under small perturbations~\cite{Howland89,Joye94,Bellissard85,
Combescure88}. In the present case, one encounters Floquet states that are 
distinctly categorized as either ``regular'' or ``chaotic'' when viewed 
semiclassically in a coarse-grained manner, but actually remain weakly coupled 
to each other on a more rigorous level, giving rise to longtime dynamics 
reminiscent of the decay of a quantum state embedded in a continuum.

The physical picture that emerges here is that of almost isolated subsystems 
of Floquet states or pre-Floquet states associated with the almost regular 
islands that are capable of sustaining longtime harmonic or subharmonic 
response to periodic driving, but eventually will decohere on timescales 
that, arguably, would considerably exceed the dynamical tunneling times of 
typical pre-Floquet states. Thus, such residual couplings might go unnoticed 
in both finite-precision numerical simulations and finite-time laboratory 
experiments.    
  
\end{appendix}

\end{document}